\documentclass[conference]{IEEEtran}
\IEEEoverridecommandlockouts
\usepackage{cite}
\usepackage{amsmath,amssymb,amsfonts}
\usepackage{algorithmic}
\usepackage{graphicx}
\usepackage{textcomp}
\usepackage{xcolor}
\usepackage{url}
\usepackage{orcidlink}
\def\BibTeX{{\rm B\kern-.05em{\sc i\kern-.025em b}\kern-.08em
    T\kern-.1667em\lower.7ex\hbox{E}\kern-.125emX}}

\makeatletter
\makeatletter
\def\ps@IEEEtitlepagestyle{%
  \def\@oddhead{}%
  \def\@oddfoot{\parbox{\textwidth}{\scriptsize \centering 
  \copyright~2025 IEEE. Personal use of this material is permitted. Permission from IEEE must be obtained for all other uses, in any current or future media, including reprinting/republishing this material for advertising or promotional purposes, creating new collective works, for resale or redistribution to servers or lists, or reuse of any copyrighted component of this work in other works. \\ \vspace{0.1cm}
  \textit{This is an extended version of the article published in the 2025 6th International Conference on Innovative Computing (ICIC). \\ 
  The final published version of record is available at: \url{https://doi.org/10.1109/ICIC68258.2025.11413235}}
  }}%
  \def\@evenfoot{}%
}
\makeatother

\begin{document}

\title{A Case for CATS: A Conductor-driven Asymmetric Transport Scheme for Semantic Prioritization\\
}

\author{
\IEEEauthorblockN{Syed Muhammad Aqdas Rizvi\,\orcidlink{0009-0004-1491-4839}}
\IEEEauthorblockA{\textit{Independent Researcher} \\
\textit{Alumnus, Lahore University of Management Sciences (LUMS)}\\
Karachi, Pakistan \\
Emails: 25100166@lums.edu.pk, s.muhammadaqdasrizvi@gmail.com}
}

\maketitle

\begin{abstract}
Standard transport protocols like TCP operate as a blind, FIFO conveyor belt for data, a model that is increasingly suboptimal for latency-sensitive and interactive applications. This paper challenges this model by introducing CATS (Conductor-driven Asymmetric Transport Scheme), a framework that provides TCP with the semantic awareness necessary to prioritize critical content. By centralizing scheduling intelligence in a transport-native ``Conductor", CATS significantly improves user-perceived performance by delivering essential data first. This architecture directly confronts a cascade of historical performance workarounds and their limitations, including the high overhead of parallel connections in HTTP/1.1, the transport-layer Head-of-Line blocking in HTTP/2, and the observed implementation heterogeneity of prioritization in HTTP/3 over QUIC. Built upon TCP BBR, our ns-3 implementation demonstrates this principle by reducing the First Contentful Paint by over 78\% in a representative webpage download configured as a deliberate worst-case scenario, with no penalty to total page load time compared to the baseline.
\end{abstract}


\section{Introduction}
Conventional transport protocols like TCP operate on a First-In, First-Out (FIFO) basis \cite{rfc793}. This semantically blind model is increasingly suboptimal for modern applications, where the timely delivery of a few critical data objects can disproportionately affect page latency and user-perceived QoE, especially on constrained networks such as 2G/3G cellular and satellite links\cite{10.1145/2964791.2901472,10.1145/2830629.2830649,10454720}. This problem is not merely theoretical, and it has practical consequences in applications ranging from web browsing and remote access (e.g., SSH/SFTP) to high-stakes tactical and 5G/6G and MR/XR networks.

To address this, we introduce \textbf{CATS (Conductor-driven Asymmetric Transport Scheme)}, a framework that modifies TCP's sending logic by introducing an active, intelligent \textbf{Conductor}. This new orchestrating logic implements an ``Asymmetric" treatment of data based on application-defined semantic importance. We chose to build this framework on TCP to leverage its decades-proven, universally deployed mechanisms for reliability, which frees our design to focus purely on scheduling and allows us to explore a practical, evolutionary path for enhancing performance across the vast ecosystem in which TCP remains the dominant transport protocol. Within the TCP framework, we specifically build upon BBR \cite{bbr-cacm17} to utilize its modern, model-based congestion control as a foundation.

While the CATS transport-native design philosophy is broadly applicable, this paper focuses its evaluation on accelerating webpage delivery. This lets us provide a clear and quantifiable evaluation using standardized QoE metrics while positioning our contribution against the well-documented evolution of HTTP performance workarounds. In doing so, CATS provides a direct response to the high overhead of parallel connections in HTTP/1.1 and the impact of transport-layer Head-of-Line (HoL) blocking in HTTP/2. Furthermore, it explores an alternative architectural trade-off to the application-managed stream model in HTTP/3 over QUIC, which does solve HoL blocking, but whose real-world performance has been shown to be inconsistent due to implementation heterogeneity \cite{Herbots2024H3PrioWild}.

The core hypothesis is that by centralizing scheduling intelligence within a TCP Conductor, we can achieve a more robust and predictable performance improvement over the default FIFO behavior. This paper's main contributions are the design of the CATS architecture with a hysteresis fairness mechanism, its implementation in ns-3, and a simulation study demonstrating a 78.7\% improvement in First Contentful Paint with negligible overhead.

\section{Related Work}

\subsection{Datacenters and Routers}
The principles of prioritization and intelligent queuing are elementary to network performance research. CATS is informed by two major areas: datacenter transport scheduling and in-network Active Queue Management (AQM).

\subsubsection{Datacenter Transport Scheduling}
The need for transport-level scheduling is most evident in datacenters, where optimizing for metrics like Flow Completion Time (FCT) is important. Earlier work, such as PDQ, established the benefits of priority and deadline-aware queuing for latency-sensitive flows \cite{Hong2012PDQ}. This was followed by influential schemes like pFabric \cite{Alizadeh2013pFabric} and PIAS \cite{Bai2017PIAS}, which presented significant performance gains by incorporating concepts such as flow size awareness and preemption, and work on PASE \cite{Munir2017PASE_TNET} explored synthesizing existing transport strategies for near-optimal performance. While these systems are tailored for datacenter traffic, they establish a strong precedent for the effectiveness of modifying transport behavior to achieve specific performance objectives through prioritization. The work herein draws inspiration from this broader willingness to evolve transport mechanisms while targeting the different problem domain of semantic importance in user-facing applications over general internet paths.

\subsubsection{Active Queue Management and Protocol Coexistence}
AQM schemes like CoDel \cite{rfc8289_CoDel} and CAKE \cite{Hoiland-Jorgensen2018CAKE} manage router queues to keep latency low by actively controlling standing queue buildup. More recent efforts such as L4S (Low Latency, Low Loss, and Scalable Throughput) extend this direction through ECN-based DualQ coupled AQM designs that aim to keep queuing delay very low even under load \cite{rfc9330}, \cite{rfc9332}. The CATS Conductor can be viewed as a source-based complement to these in-network approaches.

\subsection{QUIC and HTTP/3}
The QUIC transport protocol \cite{rfc9000}, which underpins HTTP/3, represents the most significant recent evolution in the transport layer. QUIC solves TCP’s transport-layer HoL blocking by introducing multiple independent streams, each with its own flow control, enabling highly effective application-layer prioritization schemes \cite{rfc9218}.

However, this powerful design relies on an application-managed scheduling model. This model places the burden of not only signaling priority, but also managing the complex logic of how and when to write data to these multiple streams squarely on the application. The application developer must reason about how to multiplex different resources to achieve a desired outcome. This application-level complexity is a primary source of the heterogeneity and inconsistent performance observed in real-world measurement studies, as different browsers and servers make vastly different scheduling choices even when given the same priority signals \cite{Herbots2024H3PrioWild}.

CATS explores a different architectural trade-off with a transport-native scheduling model that aims for a simplification of the application's role. In the CATS design, the application's responsibility is reduced to only providing semantic ``hints" (priorities) for its data. The complexity of scheduling, multiplexing, and fairness is not eliminated, but is instead delegated and centralized within the transport layer's Conductor. This approach has two key hypothesized advantages:
\begin{enumerate}
    \item\textbf{Predictability and Consistency:} The Conductor operates with a fixed scheduling algorithm (strict priority with hysteresis-based fairness) that is configurable at the parameter level (watermarks, payback multipliers) but invariant in its scheduling semantics. This contrasts with QUIC's application-managed model, which leaves the choice of scheduling algorithm itself to each implementation and produces the well-documented heterogeneity in real-world behaviour \cite{Herbots2024H3PrioWild}. CATS replaces algorithmic heterogeneity with parametric tunability of a fixed algorithm, which constrains the space of behaviours two CATS deployments can exhibit even when their parameter choices differ.
    \item \textbf{Integration with Transport-Layer Signals:} By operating natively within the transport layer, the Conductor has immediate access to the high-level outputs of the congestion controller, such as the calculated send budget (pacing rate and \texttt{cwnd}). This allows for highly responsive scheduling decisions that are made just-in-time, based on the most current network state available to the sender. This contrasts with application-level schedulers, which may operate on slightly delayed or coarser feedback, and it opens a rich area for future work into deeper integration with the CCA's internal model.
\end{enumerate}

Though this paper evaluates CATS within a single, coherent TCP stream, the underlying principle of a transport-native Conductor is not inherently limited. A future evolution of CATS could involve the Conductor managing a set of transport-layer streams, offering a hybrid architecture that combines QUIC's stream independence with CATS's centralized and network-aware scheduling intelligence.

\subsection{The ``Interceptor and Feeder" Architecture}
CATS is implemented using an \textbf{``Interceptor and Feeder"} model at the sender's TCP layer. This architecture gives CATS complete, segment-by-segment control over the transmission schedule.

\begin{enumerate}
    \item \textbf{The Interceptor:} The application sends data with a priority tag to the CATS socket. The \texttt{Send()} method is overridden to act as an interceptor. It does not pass the data to the base TCP buffer; instead, it places the unsegmented data into one of five internal prioritized queues and then calls the Conductor.

    \item \textbf{The Conductor:} This is the core orchestrating logic. It is designed to be re-entrant, meaning it is called repeatedly to make a fresh scheduling decision for \textit{every single segment}. It is triggered by two events: new data arriving from the application, or an incoming ACK freeing up space in the base TCP buffer.

    \item \textbf{The Feeder:} The Conductor's logic performs the feeder task. At each invocation, it inspects the prioritized queues, applies the fairness mechanism, and selects the single highest-priority eligible data chunk. It then feeds just one MSS-sized piece of this data to the base class \texttt{TcpSocketBase::Send()} method. By processing only one segment per call and then scheduling itself to run again if more data can be sent, it ensures that a high-priority packet that arrives at any time can be scheduled almost immediately, achieving a highly responsive, non-preemptive priority system.
\end{enumerate}

\begin{figure}[t!]
    \centering
    \includegraphics[width=\columnwidth]{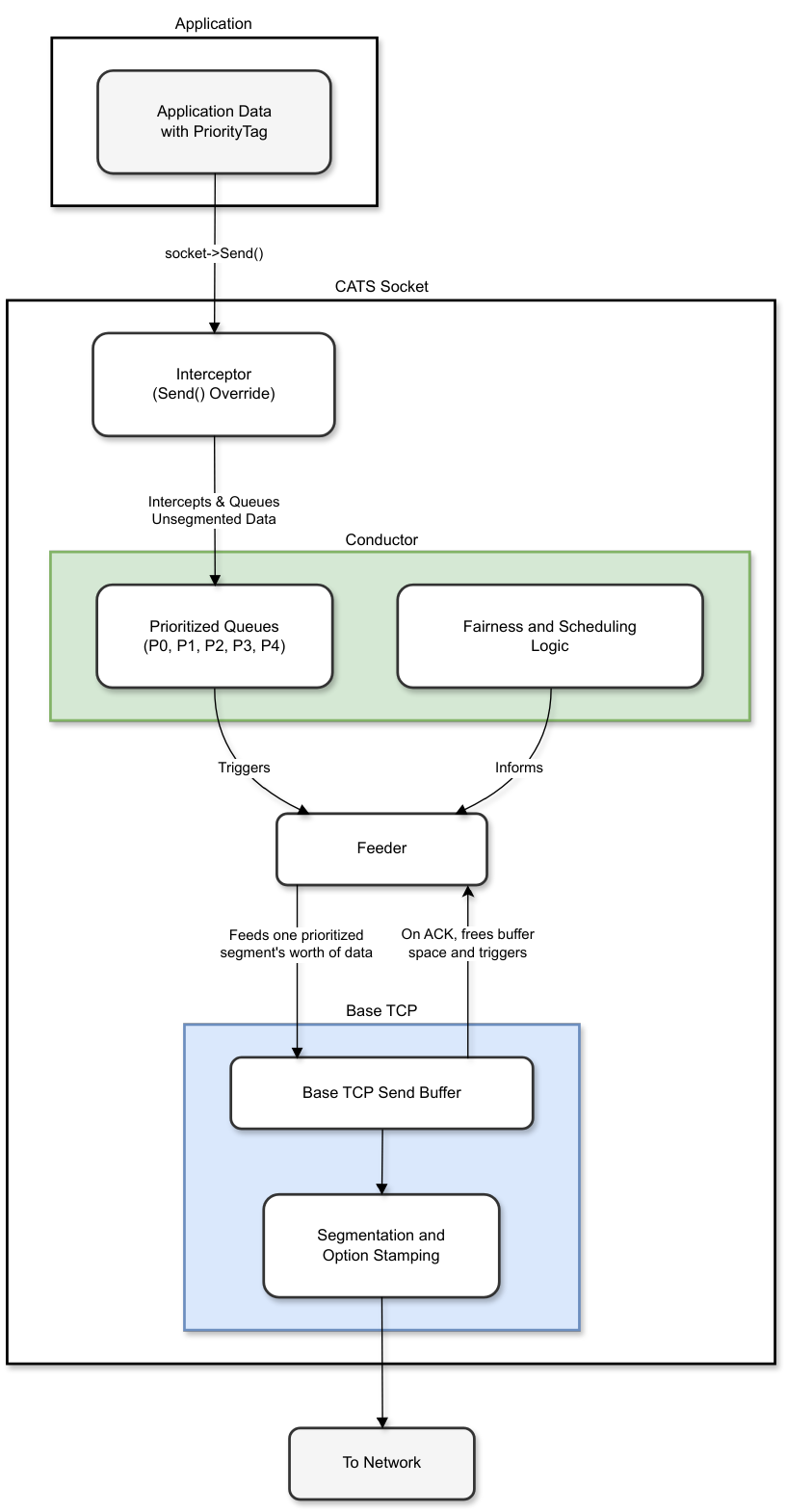}
    \caption{The CATS Interceptor and Feeder architecture. The Conductor places all intercepted application data into priority queues and it is then fed to the underlying TCP base one prioritized, segment-sized chunk at a time.}
    \label{fig:architecture}
\end{figure}

With this architecture, the CATS Conductor retains complete control over the transmission order while utilizing the proven segmentation, reliability, and congestion control logic of the underlying TCP implementation (in this case, BBR).

\subsection{The CATS API and Signaling Ecosystem}
CATS relies on a complete ecosystem for communicating priority from the application down to the wire.

\subsubsection{Priority Definition and Signaling}
Semantic priority originates at the application layer, for which a flexible, two-tiered API for communicating this to the transport layer is envisioned. A \textbf{per-message priority}, modeled in ns-3 with a \texttt{PriorityTag}, allows for fine-grained control. This would be analogous to using \texttt{sendmsg()} with a control message in a real system. A fallback \textbf{per-socket default priority}, modeled as a socket attribute, provides a simpler mechanism for applications with uniform traffic. Priorities are defined across five levels, from 0 for critical interactive data to 4 for background tasks. Although the current implementation fixes five priority levels for the webpage evaluation, the number of levels is intended as a per-use-case design parameter, such that a different application domain (e.g., three-level RTC for control/video/audio, four-level LLM serving for interactive/batch/background/telemetry) would use a different configuration.

Application developers can derive these priorities from various sources, such as parsing the standard \texttt{fetchpriority} HTML attribute \cite{mdn-fetchpriority}, using custom attributes in JavaScript frameworks, or implementing server-side heuristics at a CDN edge. As a more advanced, future direction, these priorities could be assigned automatically by Machine Learning (ML) models trained to predict the semantic importance of web resources based on DOM and CSS analysis.

\subsubsection{End-to-End Signaling via TCP Option}
While our evaluation focuses on sender-side scheduling, CATS includes a 3-byte TCP Option (using experimental Kind 254) to signal priority end-to-end. This is not only for debugging, and it enables a complete cooperative framework. The TCP Option allows for:
\begin{itemize}
    \item Receiver TCP Stack Optimization: A receiver could bypass Delayed ACKs for high-priority segments, accelerating the sender's feedback loop.
    \item Receiver Application Awareness: Priority information could be passed up to the receiving application, allowing it to prioritize processing (e.g., rendering critical CSS before executing a non-critical script).
    \item Automated QoS Cooperation: The option provides a clean hook for the sender's OS to automatically set the appropriate DSCP value in the IP header, creating a streamlined link between application intent and in-network QoS.
\end{itemize}

\subsection{The CATS Conductor and the Intra-Flow Fairness Mechanism}
The primary intelligence of CATS is the \textbf{Conductor}. The Conductor's primary role is to act as a sophisticated scheduler, enforcing the prioritization policy defined by the application. It manages the internal prioritized queues and implements a robust intra-flow fairness mechanism to prevent starvation of low-priority data.

To avoid starvation, the Conductor employs a configurable, two-threshold hysteresis model with a deadlock resolution mechanism. Let $P_i$ be a prioritized queue for $i \in \{0, 1, 2, 3, 4\}$. Each queue $P_i$ is governed by the following state variables:
\begin{itemize}
    \item $D_i(t)$: The current debt of queue $i$ at time $t$.
    \item $S_i(t)$: The state of queue $i$, either \textit{Eligible} or \textit{In-Debt}.
    \item $H_i, L_i$: The High and Low Watermark debt thresholds for queue $i$.
    \item $M_j$: The Payback Multiplier for a sending queue $P_j$.
\end{itemize}

\textbf{State Update:} After sending $N$ bytes from queue $P_j$, the state is updated. The debt of the sending queue is increased: $D_j(t+1) = D_j(t) + N$. If $D_j(t+1) \ge H_j$, the queue's state becomes \textit{In-Debt}. Simultaneously, for all higher-priority queues $P_i$ (where $i < j$), their debt is reduced: $D_i(t+1) = \max(0, D_i(t) - (N \times M_j))$. A queue becomes \textit{Eligible} again if its debt drops below its low watermark $L_i$.

\textbf{Deadlock Resolution:} A deadlock can occur if all non-empty queues enter the \textit{In-Debt} state. The Conductor resolves this via a proportional debt redistribution mechanism. Let $\mathcal{Q}_{ne}$ be the set of non-empty, in-debt queues. To ensure consistent scaling, total payback capacity is defined as a system constant: $M_{total} = \sum_{k=0}^{4} M_k$. For each queue $P_i \in \mathcal{Q}_{ne}$, its existing debt is proportionally scaled down based on its relative multiplier. 

To guarantee that a deadlocked queue $P_i$ returns to the \textit{Eligible} state natively (i.e., the new debt is strictly less than $L_i$), the configured parameters must satisfy the liveness constraint:
\begin{equation}
\frac{M_i}{M_{total}} < \frac{L_i}{H_i + \text{MSS}_{\max}}
\label{eq:parameter_constraint}
\end{equation}
where $\text{MSS}_{\max}$ represents the maximum expected segment size for the connection (e.g., 1460 bytes for standard Ethernet), bounding the maximum possible debt overshoot. Because MSS is negotiated dynamically, protocol designers must provision $H_i$ and $L_i$ against the network's maximum transmission unit. The default CATS parameters strictly adhere to this boundary condition (e.g., for $P_3$ at $\text{MSS}_{\max}=1460$B, $0.285 < 0.402$). 

To ensure absolute protocol liveness against all parameter configurations and unexpected Jumbo frames that violate Equation~\eqref{eq:parameter_constraint}, the redistribution mechanism applies an optimal bounding function. The new debt is clamped to $L_i - 1$, the supremum of the eligible state space, maximizing retained penalty while guaranteeing reactivation:
\begin{equation}
D_i(t+1) = \min \left( \text{trunc}\left( D_i(t) \times \frac{M_i}{M_{\text{total}}} \right), L_i - 1 \right)
\label{eq:deadlock_resolution}
\end{equation}
After this bounded redistribution, normal scheduling is re-applied, securely preventing infinite deadlock loops.

\section{Experimental Evaluation}
CATS was evaluated in ns-3 to validate its core behavior and quantify its performance benefits against a standard TCP BBR baseline.

\subsection{Experimental Setup}
The simulation was performed in ns-3 using a dumbbell topology with a 2 Mbps bottleneck link and a 50 ms round-trip delay. To ensure a fair and modern comparison, both the CATS and baseline scenarios use an identical, high-performance TCP stack configuration:
\begin{itemize}
    \item \textbf{Congestion Control - TCP BBR \cite{bbr-cacm17}:} BBR's model-based approach is effective on a wide range of network paths, and its paced sending provides a stable foundation for our scheduling logic to operate upon.
    \item \textbf{Loss Recovery - Proportional Rate Reduction (PRR) \cite{rfc6937}:} PRR is a state-of-the-art recovery algorithm that improves performance over standard TCP recovery, especially in the presence of bursty losses, ensuring a sturdy baseline.
    \item \textbf{RTT Estimation - TCP RTO Estimator \cite{rfc6298}:} TCP uses the standard Jacobson-Karels algorithm to maintain a smoothed RTT (\texttt{SRTT}) and RTT variation (\texttt{RTTVAR}), which are necessary for computing reliable retransmission timeouts (\texttt{RTO}) across all TCP variants. Although BBR does not use these smoothed values in its congestion-control logic, it continuously observes the raw RTT samples produced by the TCP stack to update its own independent estimate of the round-trip propagation time (\texttt{RTprop}), which is a key parameter in its bandwidth-delay product model.
    
\end{itemize}
This configuration ensures that we are evaluating our prioritization scheme on top of a strong, modern TCP baseline, isolating the impact of the CATS Conductor itself. The complete source code is openly available on GitHub\footnote{\url{https://github.com/smarizvi110/cats}}.

The sender application simulates a representative, yet challenging, webpage download. For this initial study, the experiment is designed around a single, carefully constructed webpage model. This deliberate methodological choice provides a clear and reproducible worst-case scenario that maximally stresses the transport scheduler, allowing us to isolate and analyze the primary impact of the CATS Conductor. The model's composition is based on the resource loading heuristics of modern web browsers, which internally classify resources and assign them a fetch priority (e.g., `High` for render-blocking CSS) to optimize the loading process \cite{mdn-fetchpriority}. Our five priority levels (P0=8KB HTML/CSS, P1=25KB CSS, P2=40KB JS, P3=60KB Images, P4=150KB Analytics) are a direct model of this classification, representing a realistic mix of modern web content. To create a worst-case scenario, the data groups were queued at the application layer in \textit{reverse order of priority}.

A primary difference in the implementation was the required application logic. For the baseline TCP scenario, the sender application had to implement a complex \textit{chunked sending} mechanism, inspired by ns-3's \texttt{BulkSendApplication}, to manually feed the TCP socket with small chunks of data to avoid overflowing its send buffer. For the CATS scenario, the application could simply send the entire data for all priority groups to the socket at once, as the CATS Conductor's internal prioritized queues handled the buffering and scheduling transparently, significantly simplifying the application logic.

For ensuring intra-flow fairness and preventing starvation of the low-priority bulk data during the simulation, the CATS Conductor's hysteresis parameters were configured as follows: High Watermarks ($H_{0..3}$) were set to $\{60, 30, 15, 6\}$ KB, with corresponding Low Watermarks ($L_{0..3}$) set to $50\%$ of their respective High Watermarks. The Payback Multipliers ($M_{0..4}$) were configured as $\{0.25, 0.5, 1.0, 1.5, 2.0\}$. These default values were selected to strictly satisfy the liveness constraint defined in Equation~\eqref{eq:parameter_constraint} for standard network MTUs.

\subsection{Results: Priority Group Completion}
\begin{figure}[t!]
    \centering
    \includegraphics[width=\columnwidth]{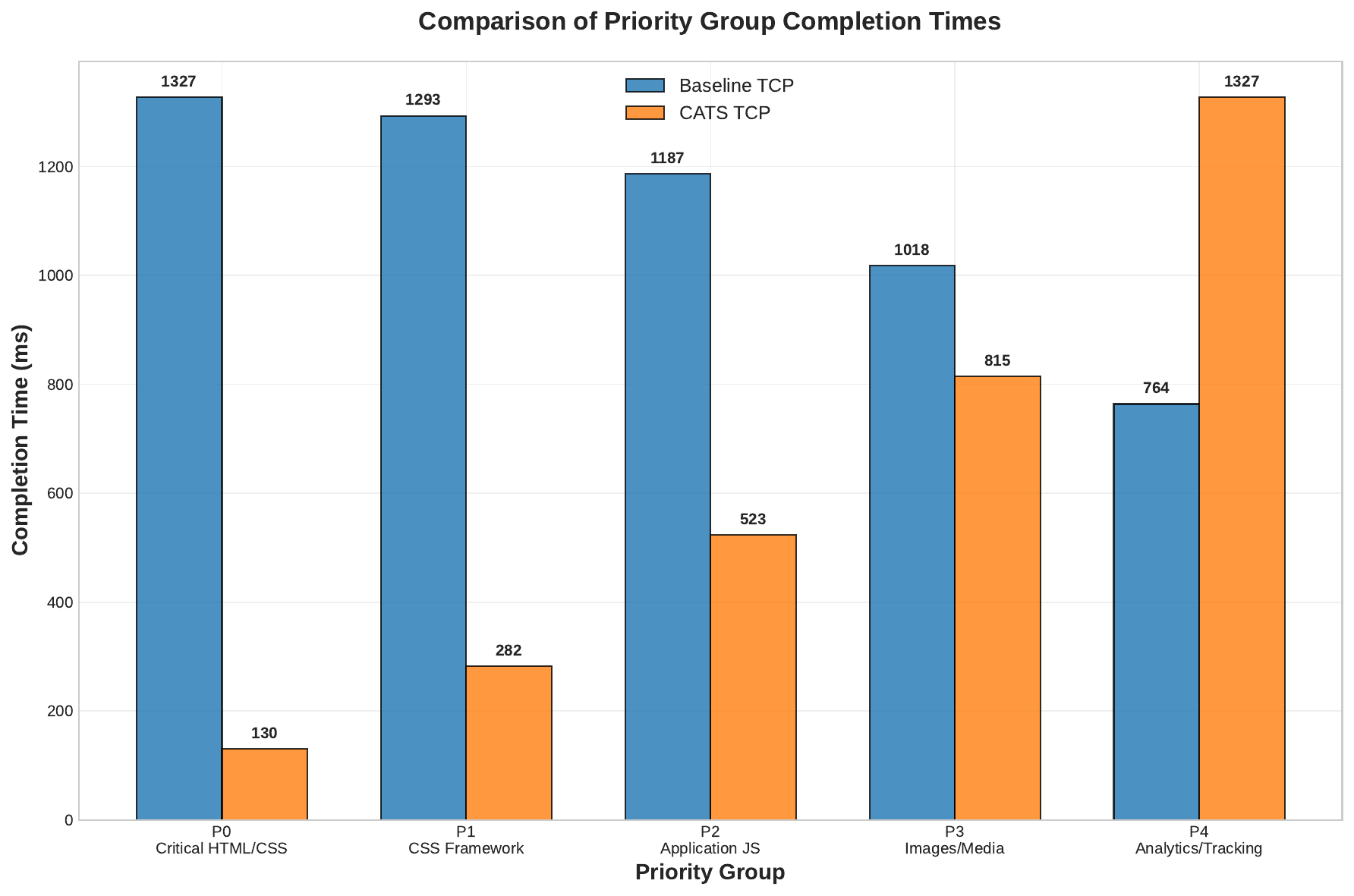}
    \caption{Comparison of Priority Group Completion Times for Baseline TCP vs. CATS, demonstrating the priority inversion achieved by CATS.}
    \label{fig:completion_plot}
\end{figure}

The main result is the completion time for each priority group. Table~\ref{tab:completion_times} and Fig.~\ref{fig:completion_plot} show that the two schemes exhibit starkly different behaviors. The baseline TCP socket adheres strictly to FIFO ordering, resulting in the most critical content (P0) being delivered last. In contrast, CATS achieves a perfect priority inversion, delivering the critical P0 and P1 content first, despite it being sent last, validating the effectiveness of the Conductor's scheduling. Additionally, Total Page Load time is identical for both scenarios (1327 ms), verifying that CATS introduces negligible overhead.

\begin{table}[t!]
\caption{Comparison of Priority Group Completion Times (ms)}
\begin{center}
\begin{tabular}{|c|c|c|c|}
\hline
\textbf{Priority}&\multicolumn{3}{|c|}{\textbf{Completion Times (ms)}} \\
\cline{2-4} 
\textbf{Group} & \textbf{\textit{Priority}}& \textbf{\textit{Baseline TCP}}& \textbf{\textit{CATS}} \\
\hline
Critical HTML/CSS (P0) & 0 (Highest) & 1327 & \textbf{130} \\
\hline
CSS Framework (P1) & 1 & 1293 & \textbf{282} \\
\hline
Application JS (P2) & 2 & 1187 & \textbf{523} \\
\hline
Images/Media (P3) & 3 & 1018 & \textbf{815} \\
\hline
Analytics/Tracking (P4) & 4 (Lowest) & \textbf{764} & 1327 \\
\hline
\textbf{Total Page Load} & --- & \textbf{1327} & \textbf{1327} \\
\hline
\end{tabular}
\label{tab:completion_times}
\end{center}
\end{table}

\subsection{Results: User Experience and Throughput}
The sharp reordering of content delivery shown in Table~\ref{tab:completion_times} has a direct and significant impact on the end-user's Quality of Experience. This can be quantified by estimating industry-standard web performance metrics from our network-level completion times, as shown in Table~\ref{tab:kpi_comparison}. It is important to note that these values represent a \textbf{lower bound} on the final render times, as they do not include client-side processing and rendering overhead. Nevertheless, they provide a clear comparison of the performance impact of the two transport schemes. CATS \textbf{yields a remarkable 78.7\% improvement in First Contentful Paint (FCP)}, demonstrating that a user sees meaningful content in less than a quarter of the time compared to the baseline.

\begin{figure}[t!]
    \centering
    \includegraphics[width=\columnwidth]{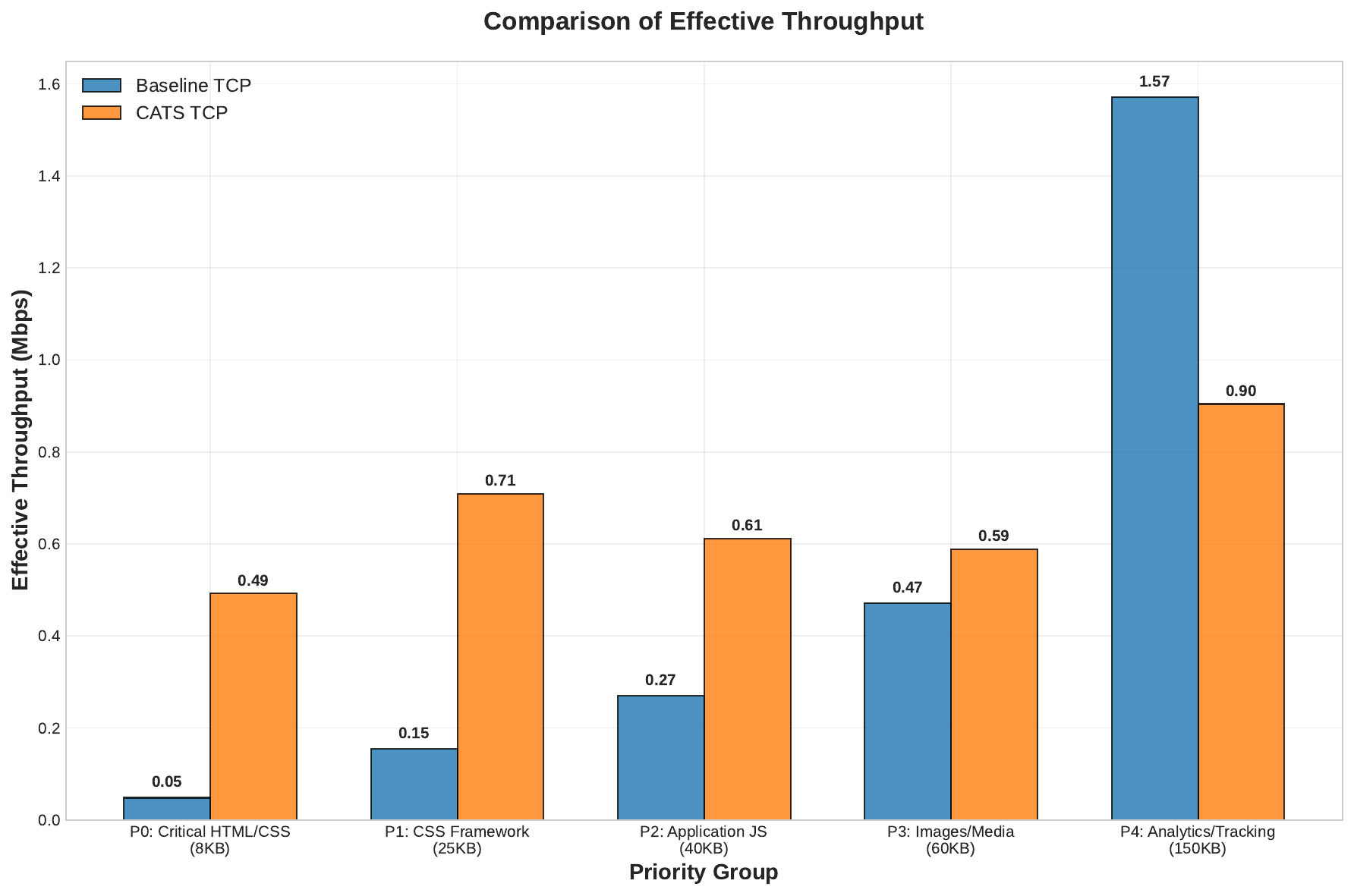}
    \caption{Comparison of Effective Throughput. Baseline prioritizes throughput for the first object sent (P4), while CATS achieves higher throughput for the prioritized objects (P0-P3).}
    \label{fig:throughput_plot}
\end{figure}

A deeper analysis of the effective throughput for each priority group reveals an interesting, non-obvious benefit of the CATS architecture, as illustrated in Fig.~\ref{fig:throughput_plot}. For the lowest-priority object (P4), the baseline TCP achieves higher throughput, as it is able to dedicate the connection's entire warm-up phase to this single large transfer. However, for all higher-priority objects (P0-P3), CATS \textbf{exhibits superior effective throughput}. This suggests that the CATS Interceptor and Feeder model keeps the core TCP send buffer lean and responsive by feeding it only one segment at a time, and is more efficient at processing and transmitting critical data chunks than the standard TCP's monolithic buffer approach. CATS correctly \textbf{prioritizes minimizing the latency for critical data} by sending it first, and its architecture concurrently \textbf{provides a more efficient transmission path for that data}, while still allowing bulk transfers to effectively utilize the bottleneck bandwidth once they are scheduled.

\begin{table}[t!]
\caption{Comparison of Estimated Web Performance Metrics}
\begin{center}
\begin{tabular}{|c|c|c|c|}
\hline
\textbf{Metric}&\multicolumn{3}{|c|}{\textbf{Performance Estimates}} \\
\cline{2-4} 
 & \textbf{\textit{Baseline TCP}}& \textbf{\textit{CATS}}& \textbf{\textit{\% Improv.}} \\
\hline
FCP$^{\mathrm{a}}$ & 1327 ms & \textbf{282 ms} & \textbf{78.7\%} \\
\hline
TTI$^{\mathrm{b}}$ & 1327 ms & \textbf{523 ms} & \textbf{60.6\%} \\
\hline
LCP$^{\mathrm{c}}$ & 1018 ms & \textbf{815 ms} & \textbf{19.9\%} \\
\hline
CLS$^{\mathrm{d}}$ & \textbf{$\le$ 0.1} & $>$ 0.1 & N/A \\
\hline
\multicolumn{4}{p{0.95\linewidth}}{\footnotesize $^{\mathrm{a}}$FCP is estimated as the completion time of the last critical rendering-path resource (P0 and P1) \cite{walton2023fcp}.} \\
\multicolumn{4}{p{0.95\linewidth}}{\footnotesize $^{\mathrm{b}}$TTI is a lab metric estimating when a page is both visually rendered and its main thread is ready to respond to user input \cite{walton2023tti}. It is estimated as the completion time of the last resource required for interactivity (P0, P1, and P2).} \\
\multicolumn{4}{p{0.95\linewidth}}{\footnotesize $^{\mathrm{c}}$LCP is a Core Web Vital \cite{walton_pollard2025lcp,mcquade_pollard2025coreVitals}. It is estimated as the completion time of our largest visual element (P3).} \\
\multicolumn{4}{p{0.95\linewidth}}{\footnotesize $^{\mathrm{d}}$CLS is a Core Web Vital measuring visual stability \cite{mihajlija_walton2023cls,mcquade_pollard2025coreVitals}. It is estimated based on the time delta between layout stabilization (P1 completion) and image rendering (P3 completion).}
\end{tabular}
\label{tab:kpi_comparison}
\end{center}
\end{table}

Finally, the differing CLS scores represent the important interaction between transport performance and application design. The baseline scenario achieves a good CLS score ($\le$ 0.1) because the image content (P3) is delivered before the layout-defining resources (P0, P1), allowing the browser to know the image dimensions before rendering. In the CATS scenario, the layout-defining content is delivered drastically faster (P1 completes at 282 ms), while the image (P3) arrives later (at 815 ms). This large delay between layout and image rendering leads to a poor estimated CLS score ($>$ 0.1). This outcome indicates that to fully capitalize on the latency benefits of an advanced transport protocol like CATS, applications must adopt modern best practices. Such practices could include reserving space for images in the initial layout via CSS/JS, or taking advantage of the CATS framework itself by sending a low-resolution, higher-priority placeholder image that is later replaced by the full-resolution, low-priority version. Essentially, an optimal user experience is the product of such cooperation between an intelligent network transport and an equally intelligent application.

\section{Conclusion and Future Work}
This paper introduced CATS, a conductor-driven asymmetric transport scheme that integrates semantic priority awareness into TCP. Our ns-3 implementation and evaluation on an elementary dumbbell topology show that CATS can dramatically improve user-perceived performance, reducing First Contentful Paint by over 78\% in a representative webpage download, with negligible overhead compared to a standard TCP BBR connection. With an empowered end-host mechanism to prioritize critical content, CATS offers a path toward a more user-responsive transport layer.

The results of this initial, exploratory study motivate several clear avenues for future research that directly address the limitations of the current evaluation.
\begin{itemize}
    \item \textbf{Advanced Simulation Scenarios and Use Cases:} To build upon the results of this paper, future work can evaluate CATS across a broader spectrum of applications and network topologies. This includes simulating video streaming workloads and analyzing the prioritization of critical video segments and modeling large file transfers (e.g., SFTP, rsync) that are interleaved with interactive control messages. These tests can be conducted on more complex topologies representative of real-world cellular and satellite networks.

    \item \textbf{High-Performance Prototyping:} For validating these findings in real-world environments, a future step can be to develop a Linux kernel prototype using the TCP Upper Layer Protocol (ULP) infrastructure and the Extended Berkeley Packet Filter (eBPF) framework. Rather than relying on an invasive and brittle kernel modification, eBPF allows the CATS "Interceptor and Feeder" architecture to be implemented safely. Specifically, \texttt{sk\_msg} programs attached to a \texttt{sockmap} can intercept application socket data before it enters the standard TCP transmit queue, allowing the Conductor logic to manage prioritization. The custom 3-byte CATS TCP Option can be dynamically injected into packet headers using the \texttt{BPF\_SOCK\_OPS\_WRITE\_HDR\_OPT\_CB} hook. Furthermore, eBPF provides direct, real-time read access to the socket's dynamically negotiated Maximum Segment Size via the \texttt{bpf\_tcp\_sock->mss\_cache} field. By storing the policy variables ($H_i$, $L_i$, and $M_i$) as multipliers in userspace-configurable BPF maps, the mathematical liveness constraint in Equation~\eqref{eq:parameter_constraint} can be calculated and enforced dynamically on a per-connection basis. This elegantly immunizes the fairness algorithm against varying network Maximum Transmission Units (MTUs) or Jumbo Frames without requiring any application-level intervention.

    \item \textbf{Application in Resource-Constrained Environments:} An interesting avenue is to evaluate CATS in the domain of the Internet of Things and embedded systems. In these environments, devices often operate with constrained hardware (limited memory, CPU), rely on a Hardware Abstraction Layer, and run on Real-Time Operating Systems, where the overhead and complexity of the network stack are primary concerns \cite{Bormann2012CoAP}. While one could argue that a multi-queue mechanism like CATS would add complexity to the transport layer, it still does put forward an architectural trade-off; the alternative is to place the burden of prioritization on each individual application, requiring developers to implement their own schedulers on top of a basic transport, which can be unpredictable, inefficient and prohibitive on constrained devices. CATS proffers a more modular and efficient model by centralizing this sophisticated scheduling and fairness logic within the transport layer itself. This would simplify application development by providing prioritization as a principal transport service, giving a more integrated and potentially more resource-efficient architecture for reliably delivering mixed-criticality data (critical sensor alerts, bulk telemetry logs, etc.) in such demanding environments.

    \item \textbf{Network Efficiency and Source-Based Load Shedding:} While the current evaluation focuses on latency reduction via data reordering, future iterations of CATS could explore active network load reduction. By integrating application-level policies (e.g., leveraging an HTTP \texttt{Save-Data: on} client hint), the Conductor could be extended to proactively discard data assigned below a certain priority threshold. Additionally, implementing source-based load shedding to drop the oldest, lowest-priority data chunks during severe, persistent congestion would preserve constrained network capacity for critical information before it even enters the network interface.

    \item \textbf{Inter-Flow Fairness and Coupled Design:} The current decoupled design is not expected to alter BBR's fundamental inter-flow fairness properties, but a coupled design, where the Conductor provides feedback to actively influence BBR's internal state machine, remains an area for exploration. Such a modification would necessitate a new, rigorous fairness evaluation against competing legacy TCP flows on a parking-lot topology.

    \item \textbf{Automated Priority Classification and Edge-Assisted Deployment:} The framework assumes that semantic priorities are explicitly provided by the application. A future direction could be to investigate automated priority classification systems that infer semantic importance without requiring manual developer annotation or application modifications. For example, ML-based classifiers deployed at CDN edges, reverse proxies, origin servers, or other network-adjacent infrastructure could analyze transmitted content and automatically assign CATS priority levels based on application-specific features such as object type, dependency structure, protocol metadata, timing sensitivity, payload characteristics, or historical interaction patterns. This could extend beyond webpage delivery to domains such as adaptive video streaming, cloud gaming, remote desktop systems, IoT telemetry, mixed-criticality industrial traffic, and AI/LLM serving workloads. Such an approach would allow semantic prioritization to be incrementally introduced into existing networked systems with minimal ecosystem disruption, enabling cooperative deployment models in which intelligence is introduced primarily at the network edge while the underlying CATS transport mechanism itself remains unchanged.
\end{itemize}

Ultimately, this research contends that by providing the transport layer with a semantic understanding of the data it carries, significant improvements in user experience can be achieved. CATS exemplifies a different angle in transport design philosophy, moving beyond a simple, blind conveyor belt for bytes towards an intelligent, user-aware orchestration system for modern applications.

\section*{Acknowledgment}
The author is grateful to Dr. Saqib Ilyas for his guidance on the approach, Dr. Syed Mohammad Irteza for his pointers on related literature, and Dr. Tariq Mumtaz for his advice on evaluation during the initial stages of this research.

\bibliographystyle{IEEEtran}
\bibliography{IEEEabrv,references}

\end{document}